 \documentclass[preprint2]{aastex}

\shorttitle{Period distribution of X-ray binaries in M31}
\shortauthors{Barnard et al.}

\begin{document}


\title{A period distribution of X-ray binaries observed in the central region of M31 with Chandra and HST}


\author{R. Barnard, J. L. Galache, and  M. R. Garcia}
\affil{Harvard-Smithsonian Center for Astrophysics, 60 Garden St, Cambridge MA 02138}
\and
\author{N. Nooraee, and P. J. Callanan}
\affil{Dublin Institute for Advanced Studies, Dublin, Ireland; University of Cork, Cork, Ireland}
\and
\author{A. Zezas}
\affil{University of Crete, Greece; Harvard-Smithsonian Center for Astrophysics}
\and
\author{S. S. Murray}
\affil{Johns Hopkins University, Baltimore, Maryland; Harvard-Smithsonian Center for Astrophysics}


\begin{abstract}
Almost all Galactic black hole binaries with low mass donor stars are transient X-ray sources; we expect most of the X-ray transients observed in external galaxies to be black hole binaries also. Obtaining period estimates for extra-galactic transients is challenging, but the resulting period distribution is an important tool for modeling the evolution history of the host galaxy. We have obtained periods, or upper limits, for 12 transients in M31, using an updated relation between the optical and X-ray luminosities. We have monitored the central region of M31 with Chandra for the last $\sim$12 years, and followed up promising transients with HST; 4$\sigma$ B magnitude limits for optical counterparts are ~26--29, depending on crowding. We obtain period estimates for each transient for both neutron star and black hole accretors. Periods range from $<$0.4 to 490$\pm$90 hours ($<$0.97 to $<$175 hrs if all are BH systems). These  M31 transients appear to be somewhat skewed towards shorter periods than the Milky Way (MW) transients; indeed, comparing the M31 and MW transients with  survival analysis techniques used to account for some data with only upper limits yield probabilities of $\sim$0.02--0.08 that the two populations are drawn from the same distribution. We also checked for a correlation between orbital period and distance from the nucleus, finding a 12\% probability of no correlation. Further observations of M31 transients will strengthen these results.
\end{abstract}





\section{Introduction}

X-ray binaries (XBs) represent the end-points of stellar evolution, and the XB population of a galaxy is informed by the star formation history of that galaxy. One way of characterizing such populations is the distribution of their orbital periods; however, the estimation orbital periods for extra-galactic XBs is extremely challenging, unless they happen to be at high enough inclinations to exhibit dips and/or eclipses. 

The most accessible extra-galactic population for period studies is the M31 X-ray transient group. These are expected to be low mass X-ray binaries  (LMXBs) with black hole (BH) primaries, since the Galactic BH LMXBs are nearly all transient \citep{mr06}; transient neutron star LMXBs also exist, but the majority of transients contain  BH primaries \citep[see e.g.][]{chen97}.  X-ray and optical observations of these transients  allow us to estimate their size, and therefore period, using the methods described below. Comparing the period distributions  of the M31 and Galactic BH transient LMXBs may shed light on possible differences in the star formation and binary evolution histories of these galaxies. 

\citet[hereafter VPM94]{vanparadijs94} found the absolute V band magnitudes of 18 Galactic LMXBs to vary between $-$5 and +5, depending on the size of the accretion disk, and X-ray luminosity. To do this, they required a sample with well known distances, and therefore limited the sample to systems  exhibiting X-ray bursts with radius expansion,  soft X-ray transients and Z-sources. The Eddington limit was assumed to be 2.5$\times 10^{38}$ erg s$^{-1}$. They found that the optical counterparts to bursters were fainter than those of non-bursters; this was a first clue that the optical and X-ray luminosities were linked, because bursters tend to have low X-ray luminosities, while the non-bursting Z-sources operate near the Eddington limit. Furthermore, previous work suggested that $M_{\rm V}$ was proportional to the orbital period \citep[e.g.][]{chevalier85,nieto90,anderson93}; 
this led VPM94 to postulate that  the  relation between X-ray and optical luminosities ($L_{\rm X}$ and $L_{\rm V}$ respectively) was caused by X-ray reprocessing in the accretion disk. 

The VPM94 toy model suggested that $L_{\rm V}$ $\propto$ $L_{\rm X}^{1/2}a$, where  $a$ is the binary separation. One might expect $L_{\rm V}$ to be proportional to  $a^2$, since that is related to the area of the disk assuming that the disc radius is proportional to $a$. However, the disk temperature  decreases with radius: $T^4$ $\propto$ $L_{\rm X}/a^2$, and $L_{\rm V}$ $\propto$ $a^2$  $T^\alpha$ where $\alpha$ $\sim$2 for the observed temperature range in LMXBs (VPM94). 

 Kepler's third  law gives $a$ $\propto$ $P^{2/3}\left(M_1+M_2\right)^{1/3}$; VPM94 simplified this by assuming that the masses of the two stars were similar, and defined $\Sigma$ as  $\left(L_{\rm X}/L_{\rm Edd}\right)^{1/2} \left(P/1\, {\rm hour}\right)^{2/3}$.  Their empirical relation
\begin{equation}
M_{\rm V} = 1.57\left(\pm0.24\right) - 2.27\left(\pm0.32\right) \log \Sigma
\end{equation}
provided a good match to observations over 3 orders of magnitude in $\Sigma$. However, their sample includes black hole and neutron star accretors, and a mix of different X-ray spectral states.

More recently \citet{russell06} examined the  correlations between optical/infra-red and X-ray luminosities for neutron star and black hole X-ray binaries separately, also factoring in the masses of the accretor and donor stars (i.e. $M_1$ and $M_2$). They found that the broadband BVRI luminosity ($L_{\rm OIR}$) correlated well with $L_{\rm X}^{0.6}$ over 8 orders of magnitude in $L_{\rm X}$ for black hole binaries, and also neutron star binaries in the hard state; the black hole normalization was $\sim$10 times higher than the neutron star normalization, consistent with the expectation that black hole systems have larger disks for a given orbital period because $\left(M_1+M_2\right)$ is larger for black holes than for neutron stars.

Moss et al. (2012, in prep, hereafter M12) focused on the $M_{\rm V}$ vs. $\Sigma$ relation for  black hole binaries in the high state at the peak of their outbursts, with the aim of providing the best period estimates for X-ray transients. They used X-ray and optical data for transients near their outburst maximum
. Their new empirical relation
\begin{equation}
M_{\rm V} = 0.84\left(\pm0.30\right) - 2.36\left(\pm0.30\right) \log \Sigma
\end{equation}
tends to result in a shorter period than the relation described in Equation 1. We note that the only significant difference between Equations 1 and 2 is in the normalization, again caused by the larger disks in the black hole systems.

We have conducted a 12 year campaign to monitor the central region of M31 for bright X-ray transients. Over 130  Chandra observations were made over this time, averaging one per month; no observations are made when M31 is too close to the sun. Bright new transients are followed up with HST ACS WFC observations using the F435W filter that approximates the B band. One observation was made shortly after the transient outburst, and another observation was made $\sim$6 months later, if there were no serendipitous coverage in an earlier observation \citep[see ][and Nooraee et al. 2012]{will05b,will05a,williams06a,barnard11a}. In this work, we present our analysis of 12 X-ray transients observed with Chandra ACIS and  HST. Herein we present the discovery of 2 new optical counterparts, and provide revised period estimates for all 12 systems.

X-ray transient behaviour appears to be governed by thermal-viscous instability in the accretion disc,  with hot and cold stable phases, along with an unstable transitional phase \citep[see e.g.][ and references within]{lasota01}. The outbursts occur when the accretion disc enters the hot stable phase; however, the resulting X-ray illumination prevents the disc from returning to the cold state until the X-ray luminosity is sufficiently diminished \citep[ hereafter KR98]{KR98}. KR98 showed that we may expect exponential decay lightcurves so long as the  the whole disc is irradiated with X-rays; otherwise, linear decay is expected. For long transient outbursts, KR98 expect the decay to switch from exponential to linear after a few  e-folding times. \citet{shah98} analyzed the decay lightcurves of various transients, and found good agreement with the theory of KR98; furthermore, they found that systems with multiple outbursts tended to exhibit the same type of decay during each outburst.

\section{Observations and data reduction}

We provide a journal of HST observations in Table~\ref{journ}. For each HST ACS observation we provide a number, the observation ID, the date, the exposure time, the pointing,  and the target (labeled T1--T12 in order of increasing RA); we also provide the closest ACIS observation, along with the time of observation with respect to the HST observation. If the transient is in a region that we have already observed, then a  dedicated ``off'' observation is unnecessary. 

\subsection{X-ray analysis}

We have examined 90  Chandra ACIS observations of the central region of M31, spaced over $\sim$12 years; these observations comprise a monitoring programme that is designed to find transient X-ray sources in the central region of M31. The  roll-angle of each observation was unconstrained, resulting in an approximately circular field of view with 20$'$ radius. We determined the position of each source from a merged 0.3--7 keV ACIS image, using the {\sc iraf} tool {\sc imcentroid}.

 The merged ACIS  image was registered to the Field 5 B band image from the Local Group Galaxy Survey observations of M31 \citep{massey06}, using  27 X-ray bright globular clusters.  We were unable to register the Chandra positions with the HST observations directly because  only a few of the HST fields contained  X-ray bright globular clusters. After refining the positions of the globular cluster in the Chandra and LGS B band  images with {\sc imcentroid}, we found the sky co-ordinates of the new positions with {\sc xy2sky}, then registered the X-ray positions to the LGS B band image using the {\sc iraf} task {\sc ccmap}. The mean r.m.s. uncertainty in position due to registration was 0.11$''$ in R.A. and 0.09$''$ in Dec.

We extracted source and background lightcurves  and spectra, along with the appropriate response files,  for each observation of all 12 sources. Source and background regions were circular, with the same radius; the radius was  optimized for crowding and off-axis angle, and ranged over $\sim$2--30$''$. The location of each background region was chosen to be near to the source region, but not so close that it might be contaminated by source flux.

 We created long term, calibrated lightcurves for each source, obtaining a 0.3--10 keV unabsorbed luminosity for each observation. We used these lightcurves to determine the overall variability of each X-ray source, and the X-ray luminosity at the times closest to each HST observation. Since we expect our transients to be black hole X-ray binaries in the high/soft state, we initially estimated the conversion from flux to luminosity by assuming a disk blackbody  emission spectrum with inner disk temperature  k$T_{\rm in}$ = 1 keV, and  $N_{\rm H}$ = 6.7$\times 10^{20}$ atom cm$^{-1}$ (the Galactic line of sight column density), then determining the unabsorbed 0.3--10 keV luminosity equivalent to 1 count s$^{-1}$. After correcting for the exposure, vignetting and background, multiplying the source intensity by this conversion factor gave the source luminosity. Source spectra with $>$200 net counts were freely fitted with disk blackbody models. When good spectral fits were found for a particular observation of a source, the parameters of these fits replaced the default parameters when estimating the source luminosity for observations of the source with $<$200 photons.

\subsection{Optical analysis}

For each ACS observation, we examined two types of images provided by the STSCI. Each observation had contained one drizzled (DRZ) image and  4--8 flat-fielded (FLT) images. The drizzled image is the final product: the flat fielded images are combined, the sky background is subtracted, and cosmic rays are removed; the image is scaled by intensity (electron s$^{-1}$). The FLT images  were corrected for instrumental effects (dark, bias, and flat); however, the sky was not subtracted, and cosmic rays were not removed; furthermore, the image was scaled to the total number of electrons received over the observation. We used the DRZ images to create the difference images described below, but the FLT files for photometry since DAOPHOT requires an accurate measurement of the sky background.

Each drizzled HST observation was registered to the LGS B band image, using the method described in the previous section. We then made a difference image for each pair of on/off observations. To achieve this, we first used the {\sc iraf} task {\sc wregister} to ensure that the pixels in each image had the same co-ordinate system and orientation. We then used the {\sc farith ftool}  to create the difference image. We examined the difference image for each observed variable X-ray source, looking for significant changes within a 3$\sigma$ radius, including the uncertainty in the centroid position, and  the r.m.s uncertainty in registration of the Chandra image to the LGS B band image; the r.m.s uncertainties in registering the HST images to the LGS image  were negligible. 

If no counterpart was found in the difference image, then we estimated the 4$\sigma$ upper limit to the B magnitude of an undetected star. To do this, we summed the counts   in each of the FLT images, within a circle with 2 pixel radius at the position of the X-ray source, for a total of $C_{\rm tot}$  counts over $T$ seconds. The maximum intensity of the counterpart is then 4$\left(C_{\rm tot}\right)^{0.5 }/ T$, which we convert to Vega B magnitude via
\begin{equation}
B = -2.5 \log\left[4\left( C_{\rm tot}\right)^{0.5}/T \right] + 25.76.
\end{equation} 
In order to ensure that our B magnitude limits were believable, we ensured that DAOPHOT was able to detect several stars with similar or fainter magnitudes in the FLT images. For images where the sky level varied significantly, we restricted the search for similarly faint stars to areas with similar sky levels as the target.

Unfortunately, we did not have the simultaneous BVRI observations to utilize the relation found by \citet{russell06}; however, we did obtain period estimates using the VPM94 and M12 methods. Both the VPM94 and M12 methods for estimating the orbital period require the absolute magnitude in the V band. We assumed that our F435W flux represented the B band flux, and used the  relation B$-$V = 0.09 that was empirically derived for LMXBs \citep{lvv01}. Systematic studies of X-ray scattering halos around Galactic LMXBs have shown that $A_{\rm V}$ $\sim$ $N_{\rm H}$ / 1.8$\times 10^{21}$ atom cm$^{-2}$ \citep{predehl95} while $A_{\rm V}$/E$\left(B-V\right)$ $\sim$3 \citep[see e.g.][]{cardelli89}; hence A$_{\rm B}$ $\sim$ $N_{\rm H} \times\left(1+1/3\right)/1.8\times10^{21}$. Finally, the distance modulus for M31 is 24.47 \citep{stanek98}. Hence
\begin{equation}\label{conv}
 M_{\rm V} = B + 0.09   -   N_{\rm H}\times\left(1+1/3\right)/1.8\times10^{21}  -    24.47. 
\end{equation}

\subsection{Estimating the periods}

When calculating the periods derived from the M12 relation, we were able to include uncertainties in the relation. We achieved this by randomly altering the $M_{\rm V}$ of every Galactic black hole binary high state in the Moss et al (2012) sample ten thousand times, drawn from a normal distribution scaled by the uncertainties, and obtaining a new $M_{\rm V}$ vs. $\log$ $\Sigma$ relation for each iteration; these $M_{\rm V}$ distributions had means and spreads that were consistent with the data. We then obtained the $\log$ $\Sigma$ corresponding to the $M_{\rm V}$ of each target in every iteration, and derived the mean log $\Sigma$ and uncertainties for each target by fitting a Gaussian to the resulting log $\Sigma$ distribution. Unfortunately, we did not have access to sufficient information to include uncertainties in the VPM94 relation; instead, we propagated the uncertainties in $M_{\rm V}$ and $L_{\rm X}$ assuming that the relation was fixed.

\section{Results}
\label{res}

\begin{figure*}
\epsscale{2.2}
\plotone{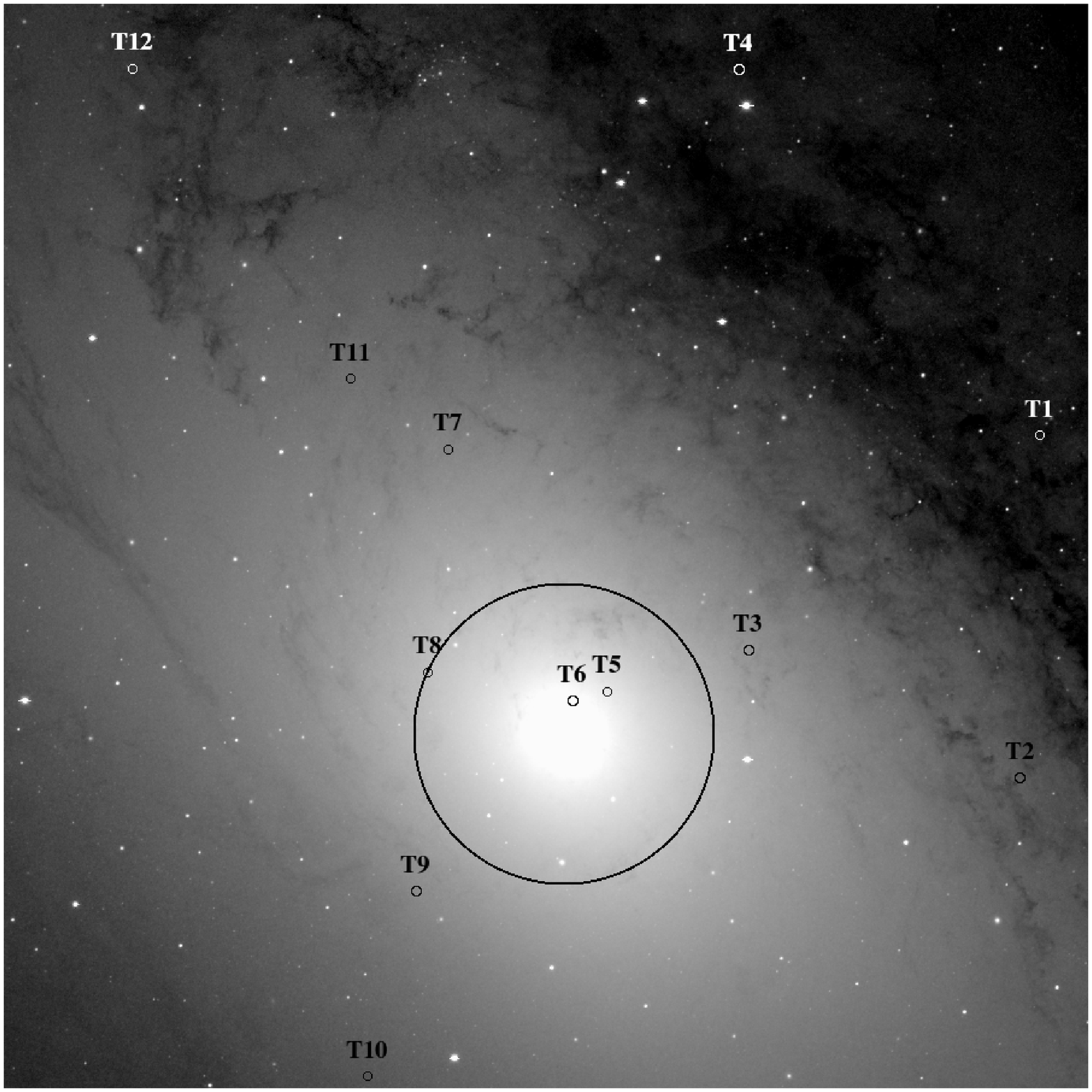}
\caption{Locations of the 12 transients, superposed on a B band image from the LGS (Massey et al., 2006). The area enclosed by the circle is within 100$''$ of the nucleus; this region is thought to produce X-ray binaries at an enhanced rate, due to its high density \citet{voss08}.} \label{btrans}
\end{figure*}

Figure~\ref{btrans} shows the locations of the 12 transients on a log-scaled B-band image of the central region of M31, provided by the LGS (Massey et al., 2006).  Table~\ref{translist} gives the positions of each target, and summarizes the results discussed later in this section; we provide the 0.3--10 keV luminosity during the closest ACIS observation to the HST on observation, our estimated 0.3--10 keV luminosity at the time of the HST on observation, the B magnitude (or 4 $\sigma$ upper limit) for the counterpart, and period estimates using the empirical relations of VPM94  and M12. We present the $\sim$12 year, 0.3--10 keV unabsorbed  luminosity  lightcurves of the 12 transients in Fig.~\ref{tran1}.

\subsection{Analysis of the Chandra observations}

Only four of the targets had $>$200 photons in the Chandra observation that was closest to the HST on exposure: T1, T2, T5, and T9.  Spectral fit parameters are provided in Table~\ref{spectab} for absorbed disk blackbody models. The T9 spectrum was piled up; we account for this in Noorae et al. (2012, submitted). Only T1 apparently experienced greater than Galactic absorption, with $N_{\rm H}$ = 2.5$\pm$1.9$\times 10^{21}$ atom cm$^{-2}$ (uncertainties are quoted at the 90\% confidence level). We find that  $A_{\rm B}$ = 1.9$\pm$1.4 for T1, and 0.50 for the remaining transients. 

 VPM94 obtained  X-ray luminosities from the Ariel V X-ray catalog, and Ariel V  employed the 2--10 keV band \citep{warwick81}. Hence we estimated the 2--10 keV luminosity of each target at the time of the HST observation. For those transients with freely fitted spectra, we used the $L_{2-10} / L_{0.3-10}$ factors provided in Table~\ref{spectab}.  For transients with no freely fitted spectra, we assume  k$T_{\rm in}$ = 1 keV, giving $L_{2-10} / L_{0.3-10}$ = 0.48.

 However, it is possible that certain transients were in their low/hard state at the time of the HST on observation; the transition from low state to high state during the rise of the outburst occurs at a luminosity $\la$10\% Eddington \citep{glad07,tang11}, while the transition from high state to low state during the decay occurs at $\sim$2\% Eddington due to hysteresis \citep[see e.g.][ and references within]{maccarone03}.  The low state may be characterized by a power law with photon index $\Gamma$ $\sim$ 1.7; assuming a power law with $\Gamma$ = 1.7 rather than a disk blackbody with k$T_{\rm in}$ = 1 keV results in 2--10 keV luminosities that are $\sim$75\% higher, and periods that are $\sim$35\% shorter. Since 8 of the 12 transients only yielded upper limits to their periods, we present the more conservative upper limits obtained by assuming high state spectra. 

\subsection{Analysis of the HST observations}
We present the difference images for T1--T12 in Fig.~\ref{difim}, showing the difference in intensity between the drizzled images of  the  on and off observations; white is positive, and black is negative. The 1$\sigma$ and 3$\sigma$ uncertainty ellipses are shown, including uncertainties in X-ray position, and X-ray registration to the LGS B image. North is up, east is left. Each image shows a 5$''$$\times$5$''$ field. 
The background level varied considerably, with the nuclear region being drastically more crowded than the outer regions; as a result, 4$\sigma$ detection limits ranged over B$\sim$26--29.

\subsection{Discussion of individual sources}

\subsubsection{T1}

Chandra, Swift, HST and Einstein observations of T1 are discussed in detail by \citet{barnard11a}. T1 has no ACS off image with the F435W filter, because the WFC was not in use at the time. Instead, we used a serendipitous 81 s V band ACS/WFC observation (ID j92ga5naq, 2005-02-17, PI Crotts). There are many black stars in the difference image, because they are brighter in the V band than in the B band; however, there are no white stars to signify the transient, and we estimate that B $>$28.7. The sky background was extremely low, and we detected  several stars over the image with B $\geq$28.7 at the 4$\sigma$ level in one FLT image. Since there was no B band off image, we examined the 8 individual FLT images to determine whether there were any detections of stars within the 3$\sigma$ ellipse; this allowed us to discriminate between stars and e.g. cosmic rays. We found no stellar candidates that were detected in $>$3 out of 8 FLT images, and conclude that there were no bright B band stars within the 3$\sigma$ ellipse.

For T1, the closest ACIS observation to the HST on image was  Obs 7139 (day $\sim$2500  in Fig. 2),  taken 27 days previously; however, T1 was not covered by  any ACIS observations for $>$100 days either side of the 2006 outburst. In a later outburst (day $\sim$3900), the 0.3--10 keV luminosity went from 8.6$\times$10$^{37}$ to 0.4$\times$10$^{37}$ erg s$^{-1}$ over 27 days. Assuming exponential decay, the e-folding time was 8.8 days. We applied the same exponential decay to the 2006 outburst, and found a 0.3--10 keV luminosity of 0.5$\times$10$^{37}$ erg s$^{-1}$ at the time of the HST on observation. For the peak spectrum, we obtained a best fit k$T_{\rm in}$ of 1.5$\pm$0.3 keV; as a result, $L_{2-10} / L_{0.3-10}$ = 0.55. 

\subsubsection{T2}

The T2 outburst (day $\sim$2550) lasted $\sim$150 days. The X-ray peak was during Obs 7140, 32 days prior to the HST on observation.  Fitting the lightcurve with exponential decay yielded an e-folding time of 38 days; however, the residuals were large, and   the $\chi^2$/dof was unacceptable (47/3). We also fitted the lightcurve with power law and straight line models; the best fit values had worse $\chi^2$/dof than the exponential fit. There was insufficient data to estimate the intensity for a  model where the decay went from exponential to linear, hence the exponential decay model provided the best estimate of the 0.3--10 keV luminosity of T2 at the time of the HST on observation. Using the best fit e-folding time of $\sim$40 days, we obtained  a 0.3--10 keV luminosity of  $\sim$6$\times$10$^{37}$ erg s$^{-1}$ for the HST on observation. The best fit disk blackbody model for the peak spectrum yielded k$T_{\rm in}$ = 0.54 $\pm$0.04 keV; this results in a 2--10 keV luminosity of just 1.5$\times 10^{37}$ erg s$^{-1}$. 

The positioning of the T2 on observation was off by $>$1$'$, due to the misidentification of the guide star. As a result, it was impossible to accurately register the FLT images to the drizzled image, and  this limited our photometric accuracy. The difference image shows one particularly variable star; its B magnitude changed from $\sim$24.7 to $\sim$25.6 between the on and off observations. We therefore consider this star to be the most likely counterpart. Since the our optical counterpart to T2 had a  B magnitude $\sim$25.6 even after the X-rays had disappeared, we removed this component when calculating the period. 

\subsubsection{T3}
T3 was first studied by \citet{will05}, who found no counterpart and concluded that B $>$25.5. However, our method of estimating the magnitude limit for the counterpart yielded B $>$27.2. 

T3 exhibited two outbursts, $\sim$2000 days apart; the HST observations were triggered by the first outburst. This  outburst had an observed peak 0.3--10 keV luminosity of 3.7$\pm$0.3$\times$10$^{37}$ erg s$^{-1}$, 22 days before the HST on observation (day $\sim$1700); however, the next observation of T3 was $\sim$150 days afterward, where the 0.3-10 keV luminosity had fallen to 0.35$\pm$0.09 $\times$10$^{37}$ erg s$^{-1}$. If we assume that these observations cover a single exponential decay, then the e-folding time is 68 days. The 0.3--10 keV luminosity would then be $\sim$2.73$\times$10$^{37}$ erg s$^{-1}$ at the time of the HST on observation.

\subsubsection{T4}

T4 was mostly quiescent, but exhibited a $\sim$1500 day period of X-ray  activity, starting on day $\sim$1500 where the intensity varied by a factor $\sim$20.  The HST on observation was triggered at day $\sim$2500.  The 0.3--10 keV luminosity of T4 was $\ga$1.5$\times 10^{37}$ erg s$^{-1}$ for the Chandra observations closest to the HST on observation, and $\sim$0.4$\times 10^{37}$ erg s$^{-1}$ at the time of the off observation.

We found no evidence for stellar variability in the difference image of T4; we set $B$ $>$28.9.

\subsubsection{T5}

T5 was first discussed by \citet{williams06a}, who found no counterpart, and set a magnitude limit of B $>$ 24.7 due to the high sky background caused  by proximity to the nuclear region. Our careful analysis of the FLT images from this observation with DAOPHOT improved the magnitude limit to B $>$ 26.6.

T5 exhibited a single outburst that was observed for $\sim$80 days, starting at day $\sim$1750. Exponential fits to the decay lightcurve were rejected ($\chi^2$/dof $\ge$63/3); however the lightcurve was well described by linear decay with slope $-$0.06 ($\chi^2$/dof = 3.5/3). The HST on observation occurred 29 days after the observed peak, hence we estimate the 0.3--10 keV luminosity to be 4.0$\times 10^{37}$ erg s$^{-1}$. The brightest observation yielded a spectrum with k$T_{\rm in}$ = 1.1$\pm$0.2, yielding a 2--10 keV luminosity of 2.5$\times 10^{37}$ erg s$^{-1}$.

\subsubsection{T6}

We discovered from our long-term ACIS lightcurve of T6 that it is not a transient,  but does vary by a factor $\ga$10. Fortunately, T6 was observed 6 times in our survey, and we were able to make a difference image from the two observations with the highest  contrast in X-ray luminosity (Observations 3 and 4 in Table 1). We found no evidence for variability in the difference image within the 3$\sigma$ ellipse; instead, we estimate B to be $>$26.3.

  We estimate that the 0.3--10 keV luminosity of T6 varied by a factor $\sim$4 between  HST observations 3 and 4 (as described in Table 1), from $\sim$0.9$\times$10$^{37}$ to $\sim$0.2$\times$10$^{37}$ erg s$^{-1}$.

\subsubsection{T7}

T7 exhibited a slow, faint outburst that lasted $\sim$250 days, starting at day $\sim$2300. The shape of the outburst was unlike the classical fast rise followed by exponential decay  (FRED); instead, the lightcurve was approximately flat for the four observations either side of the HST on image, with a 0.3--10 keV luminosity of $\sim$ 0.6$\times$10$^{37}$ erg s$^{-1}$.

Since the 0.3--10 keV luminosity of T7 did not significantly vary between its on and off observations, we cannot expect to detect optical variability in the counterpart. As a result, we must set our magnitude limit to that of the brightest star within the 3$\sigma$ position uncertainty ellipse, giving B $>$25.0.  

\subsubsection{T8}

T8 was initially quiescent, turned on around day 1400, and has been active ever since. The 0.3--10 keV luminosity was $\sim$0.4$\times$10$^{37}$ erg s$^{-1}$ at the time of the HST on observation, and also at the the time of the off observation. 

As with T7, there was no significant difference between the X-ray luminosity of T8 at the times of the HST on and off observations. Hence we set our magnitude limit to be fainter than the brightest star within 3$\sigma$ of the X-ray position, i.e. B  $>$23.9. 

\subsubsection{T9}
T9 was brightest X-ray source ever seen in M31, and is discussed in detail in \citet{kaur12} and  Noorae et al. (2012). The difference image shows a clear counterpart within the 1$\sigma$ region.  B = 23.91$\pm$0.08. No counterpart was observed in the ``off'' image, and  B $>$27.8 at the 4$\sigma$ level; hence, we assume that all of the optical light comes from reprocessed X-rays.

 The closest ACIS observation was 14 days after the  HST on image. However, the observed peak occurred in a HRC observation, 34 days before the HST on observation. The decay lightcurve appears to transition from exponential decay to linear decay (see Nooraee et al., 2012); we estimate the 0.3--10 keV luminosity to have been $\sim$50$\times 10^{37}$ erg s$^{-1}$ at the time of the HST on observation. Using a disk blackbody emission model, and correcting for pile-up effects, yielded k$T$ $\sim$0.8 keV; as a result, we estimate the 2--10 keV luminosity to have been $\sim$18.6$\times 10^{37}$ erg s$^{-1}$ at the time of the HST on observation.

\subsubsection{T10}

T10 is discussed in detail by \citet{will05b}, who used DAOPHOT and ALLSTAR to disentangle the counterpart from its bright neighbor. The counterpart dominated the blend for the on observation, but was significantly fainter in the off observation, allowing \citet{will05b} to obtain B = 24.90 for the other star. Assuming the magnitude of the other star was 24.90 during the on observation yielded B = 24.52$\pm$0.08 for the counterpart during the on observation, and 24.95$\pm$0.08 during the off observation. As with T2, we subtracted the ``off'' component of the optical emission when calculating the orbital period for T10.

The T10 outburst was captured in 5 ACIS observations over $\sim$180 days; the shape is flat-topped, rather than the typical fast rise followed by exponential decay. The closest ACIS observation occurred 1 day before the HST on observation, with 0.3--10 keV luminosity 3.0$\pm$0.2$\times$10$^{37}$ erg s$^{-1}$; given the slow evolution of the outburst, we assume the same luminosity at the time of the HST on observation.

\subsubsection{T11}
The T11 outburst was significantly detected in 5 ACIS observations over 54 days. The decay was well described by an exponential with an e-folding time of 33 days ($\chi^2$/dof = 2/5). We calculated the 0.3--10 keV luminosity to be $\sim$0.34$\times$10$^{37}$ erg s$^{-1}$. 

The T11 difference image revealed a likely  counterpart within the 1$\sigma$ ellipse. We found that B =24.87$\pm$0.09 for the on observation. We found no trace of a counterpart in the ``off'' observation; B $>$28.2 at the 4$\sigma$ level. Hence, we assume that all of the optical light comes from reprocessed X-rays. 

\subsubsection{T12}

T12 was first discussed by \citet{will05}, who identify a counterpart despite the fact that it is brighter in the off observation than in the on observation; it appears as a black star in Fig.~\ref{difim}. We disagree with this interpretation, since there is no scenario where the disc brightens during decay, and instead impose a 4$\sigma$ upper limit of B $>$ 27.9.

T12 exhibited an apparently double-peaked outburst over $\sim$200 days, with the second peak brighter than the first.  However, the previous observation to include T12 was $\sim$150 days earlier, meaning that the true peak was not observed. The 0.3--10 keV luminosity fell from 2.4$\pm$0.3$\times 10^{37}$ to 0.26$\pm$0.09$\times 10^{37}$ erg s$^{-1}$ over 61.5 days; if we assume exponential decay, then the 0.3--10 keV luminosity was $\sim$0.9$\times 10^{37}$ erg s$^{-1}$ at the time of the HST on observation.

\subsection{Estimating the periods for the 12 targets}

Since the e-folding time for transient decay is $\sim$2.2 times longer in the optical than in the X-ray lightcurves, the VPM94 and M12 relations only truly apply to the peak of the outburst, and get more inaccurate as the burst evolves. We made two sets of period calculations for each source: one set  estimating the X-ray luminosity of each transient at the time of the HST observation, and one set estimating the optical luminosity at the time of the X-ray peak. Comparison of the two sets of periods indicated the robustness of these estimates. 

For T1, the NS periods agreed within to within 30\%, and the BH periods varied by 40\%; the e-folding time was particularly short for this system.  For T9  there was a $\sim$30\% discrepancy between the  periods.  For T12, the  periods differed by a factor $\sim$3; this large discrepancy may be due to the double-peaked nature of the T12 outburst. The two sets of periods agreed to within $\sim$10\% for all targets not previously specified. These results suggest that the observed peaks were generally close enough to the true peaks for the VPM94 and M12 relations to hold.

The periods listed in Table~\ref{translist} are those obtained by estimating the X-ray luminosity at the time of the HST on observation; we provide 1$\sigma$ uncertainties for systems with identified counterparts, and 3$\sigma$ upper limits to the periods for systems with magnitude limits.  We note that some of the transients are rather X-ray faint, suggesting that they are more likely to be in the low/hard state, rather than the thermal high state; the bolometric luminosities of these sources would then be somewhat higher, leading to shorter periods. However, most of these systems only have upper limits to the period already; T11 is the only X-ray faint source with a counterpart.

We note that \citet{will05} fitted the ACIS spectra of T10 despite their low quality ($\sim$120 net source counts). They favor power law emission models with N$_{\rm H}$ $\sim$2$\times 10^{21}$ atom cm$^{-2}$; they dismissed disk blackbody emission models because the $N_{\rm H}$ was below the line-of-sight value instead of fixing $N_{\rm H}$. As a result, they found L$_{\rm X}$ = 6$\times 10^{37}$ erg s$^{-1}$, and $M_{\rm V}$ = $-$1.3$\pm$0.5, both a factor $\sim$2 brighter than our values (see Table~\ref{translist}); these results give a $\sim$130 hr period, considerably longer than the $\sim$50 hr period that we obtained. However, we checked these spectra and found $N_{\rm H}$ to be consistent with the Galactic line-of-sight absorption, and we prefer the results from our assumed spectral model.

\subsubsection{Comparison of  the M31 and MW black hole transient  periods}

In Fig.~\ref{perdist} we order the transients by increasing  BH period; the y axis serves to separate them, but does not reflect any physical parameter. Circles and stars represent NS and BH periods respectively, for systems with counterparts; otherwise, we give the 3$\sigma$ upper limits, represented by arrows. We show the period distribution of 15  Milky Way BH binaries  listed in \citet{mr06} for comparison, represented by squares; the three persistently bright BH HMXBs are excluded. Our period distribution for the M31 transients is similar to that of the Galactic BH LMXBs; however the M31 transients may tend towards shorter periods.

We have not included the period for T8 since its X-ray magnitude did not vary between the HST on and off observations, meaning that our B band magnitude limit was set by the brightest star within the 3$\sigma$ positional uncertainty. This limit is brighter than most of the observed counterparts, and the resulting period limit is not useful. While the X-ray luminosity of T7 did not vary  between the HST on and off observations either, the B magnitude limit was fainter than any of the observed counterparts so we have included T7.


We tested the possibility that  the periods of the transients in M31 are consistent with 
those in the Milky Way using the  tests for univariate data implemented in the ASURV package; this method allows us to compare a censored population, containing some systems with upper limits only, with an uncensored population, where all the values are known, within uncertainties \citep[see e.g.][]{peto72,lavalley92}. We find that the two 
distributions are inconsistent at the $\sim$92--98\% confidence level.
In order to assess the effect of uncertainties on the orbital periods of 
the M31 systems we repeated the test several times by randomly sampling 
periods assuming  that they follow a Normal distribution with the mean 
and standard deviation calculated from the values in Table 2. In all 
cases the results give very similar confidence levels. Our  sample of M31 transients  differs from the MW transients at the $\sim$2$\sigma$ level; increasing the M31 transient population should amplify this result. 

\subsubsection{Trends in period and B magnitude limit vs distance from the M31 nucleus}

\citet{voss08} discovered an enhanced population of X-ray binaries within $\sim$100$''$ of the M31 nucleus. The radial distribution of ``primordial'', i.e. expected, LMXBs follows the distribution of stellar mass in the galaxy \citep{gilfanov04}; however, the surplus population appears to follow the $\rho_{*}^2$ distribution expected of dynamically formed binaries \citep{fabian75}. Dynamically formed LMXBs are normally very rare outside globular clusters; however,  \citet{voss08} propose that the bulge of M31 is sufficiently large and dense to allow the formation of a significant number of dynamically formed binaries. However, the stellar velocities in the M31 bulge are $\sim$5--10 times higher than in globular clusters; hence \citet{voss08} expect only short period binaries to survive, with most of the survivors containing black holes.

Since the number of dynamically formed black binaries is strongly dependent on the distance from the center ($r_{\rm gc}$), one might expect to observe some correlation between orbital period and radial distance for our transients.
 In the top panel of  Fig.~\ref{rvp} we plot $r_{\rm gc}$ vs. $P$ for each transient, excluding T8. The bottom panel of Fig.~\ref{rvp} shows F435W ($\sim$B band) detection limit for each transient; T7 lies in a diagonally-shaded region because it  showed no X-ray variability between the HST on and off observations, hence the B magnitude was limited by the brightest star within the 3$\sigma$ location ellipse rather than the local background. The sensitivity increased steadily with increasing $r_{\rm gc}$,    varying by 2.5 magnitudes (i.e. a factor 10).

 We employed the Cox hazard test and the Kendal tau test, both implemented in ASURV to deal with bivariate data, to compare the population located 20$''$ $<$ $r_{\rm gc}$ $<$150$''$  with the population at 150$''$ $<$ $r_{\rm gc}$ $<$350$''$; the probability that these two populations were drawn from the same population  ($p_0$) is $\sim$0.12.  Limiting $r_{\rm gc}$ to 300$''$, thereby removing T2, and re-running the Cox hazard and Kendal tau tests produces the same result.  By contrast, extending $r_{\rm gc}$ to 400$''$ for the second group, thereby including T1,  eradicates any trace of a correlation. Therefore, a more substantial transient population is required  to confirm this suspected correlation.

\subsubsection{Decay profile vs. X-ray luminosity and  estimated period}

 X-ray transients are expected to exhibit exponential or linear decay, depending on whether or not the disc is fully ionized; this depends on the orbital period and peak X-ray luminosity  \citep{shah98}. Hence we used Fig. 1 of that paper  to determine whether the decays of our targets should be exponential or linear.  

With two exceptions, all of our targets are expected to exhibit exponential decay; T7 is consistent with either linear or exponential decay, while T11 should decay linearly according to \citep{shah98}.

 Interestingly the decay of T11 was  definitely more  exponential ($\chi^2$/dof = 2/5)  than linear ($\chi^2$/dof = 93/5); we suggest that the true peak luminosity of T11 was considerably higher than the observed peak 0.3--10 keV peak of $\sim$4$\times$10$^{36}$ erg s$^{-1}$. The true orbital period of T11 could then be significantly  shorter than  our estimate of 150$\pm$30 hr.

 Furthermore, the T5 outburst appears to have decayed linearly, even though exponential decay is expected; no observations were made in the $\sim$170 days prior to the observed outburst, hence we may have caught the linear tail of a longer and brighter outburst. A brighter peak X-ray luminosity for T5 would mean a shorter period; however, this would not affect our results, since we only have an upper limit for T5.


\section{Summary and conclusions}

We report new orbital period estimates for 12 X-ray transients observed by Chandra and HST. We have identified counterparts to 4 X-ray sources, 2 of those are new to this work; we have estimated 4$\sigma$ upper limits to the B magnitudes of the remaining 8 systems. Period estimates were previously published for 5 sources (T1, T3, T5, T10, T12) , using the VPM94 empirical relation \citep{will05b,will05a,will05,williams06a,barnard11a}, and Noorae et al. (2012) have already applied the M12 relation to T9. 
We obtained periods that ranged from $<$0.17 hours  to $<$175 hours for black hole LMXBs using the M12 relation, and $<$0.4--490$\pm$90 hours for neutron star LMXBs using the VPM94 relation. 

The M31 transients  that we have followed up appear to favor shorter periods than the Galactic black hole transients; if all of our transients  contain BHs, then the probability for the M31 and MW binaries being drawn from the same population is $<$0.08. This result may be explained by an excess of X-ray binaries dynamically formed in the high density bulge of M31; such systems require short periods to survive \citep{voss08}. However, some of the transients with the shortest periods may contain neutron stars.

We searched for a correlation between the period and distance from the nucleus that would further support the hypothesis of \citet{voss08}. We found a hint of a correlation, at the $\sim$90\% confidence level.

Further observations of M31 transients with Chandra and HST will add significance to these  relations.



\section*{Acknowledgments}
 We thank the referee for their thoughtful comments that helped clarify this work. This research has made use of data obtained from the Chandra data archive, and software provided by the Chandra X-ray Center (CXC).
R.B. is funded by Chandra grant GO9-0100X and HST grant GO-11013. M.R.G. and S.S.M are both  partially supported by NASA contract NAS8-03060.



{\it Facilities:} \facility{CXO (ACIS)} \facility{HST (ACS)}.



\bibliographystyle{aa}


\clearpage



\begin{figure*}
\epsscale{2.2}
\plotone{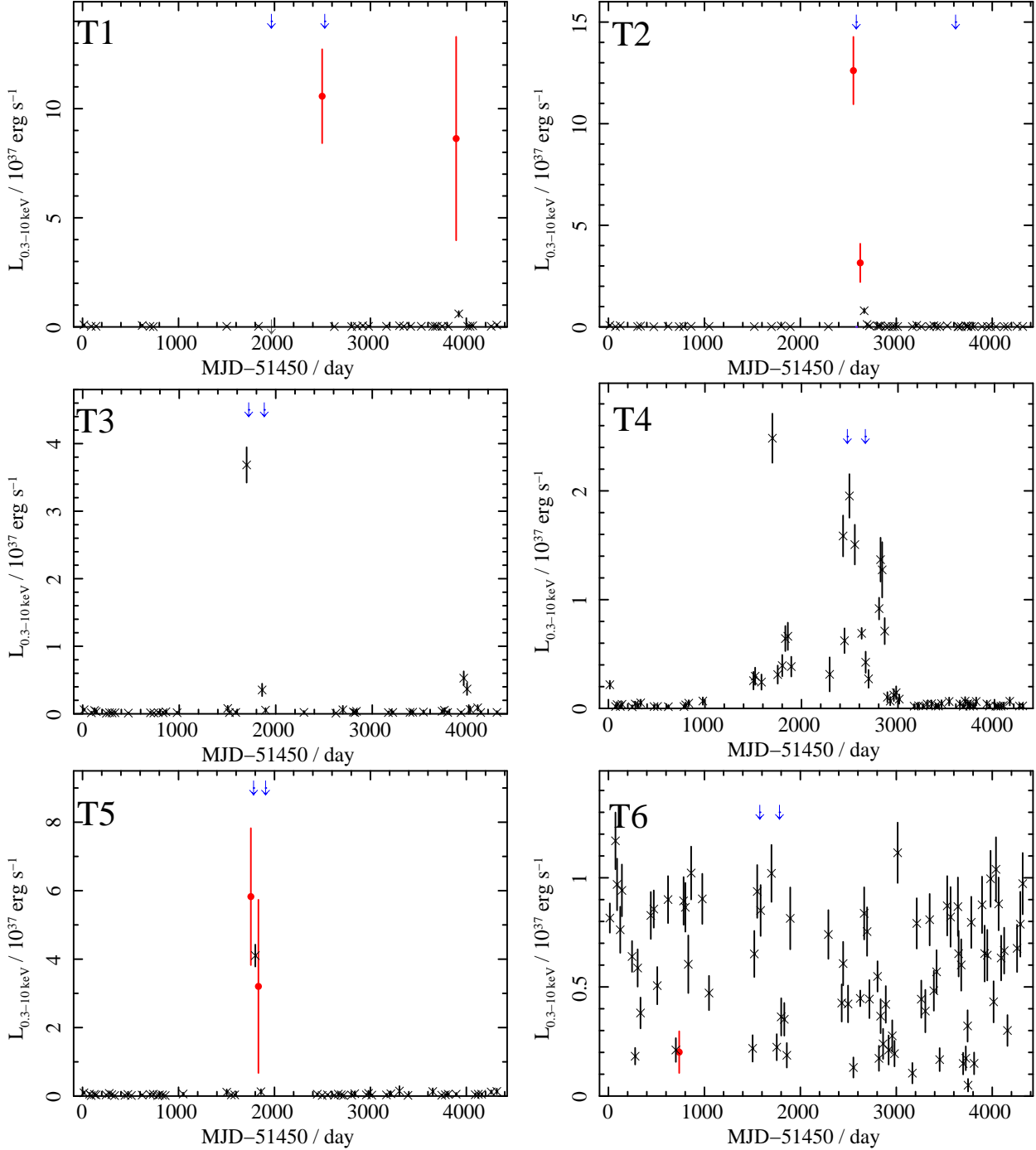}
\caption{Long-term ACIS  0.3--10 keV luminosity lightcurves of each transient. Red circles indicate observations with $>$200 net source counts; their spectra were freely fitted with absorbed disk blackbody models, and their uncertainties include uncertainties in the model parameters. Black crosses indicated observations with $<$200 net source counts; the luminosities for these observations assume spectral parameters equal to the best fit constant values for the freely-fitted observations; uncertainties for these observations are derived directly from the intensity uncertainties. The times of the HST observations are indicated by blue downward-facing arrows; the ``luminosities'' of these points were chosen for clarity, and have no physical meaning.  }\label{tran1}
\end{figure*}

\addtocounter{figure}{-1}
\begin{figure*}
\epsscale{2.2}
\plotone{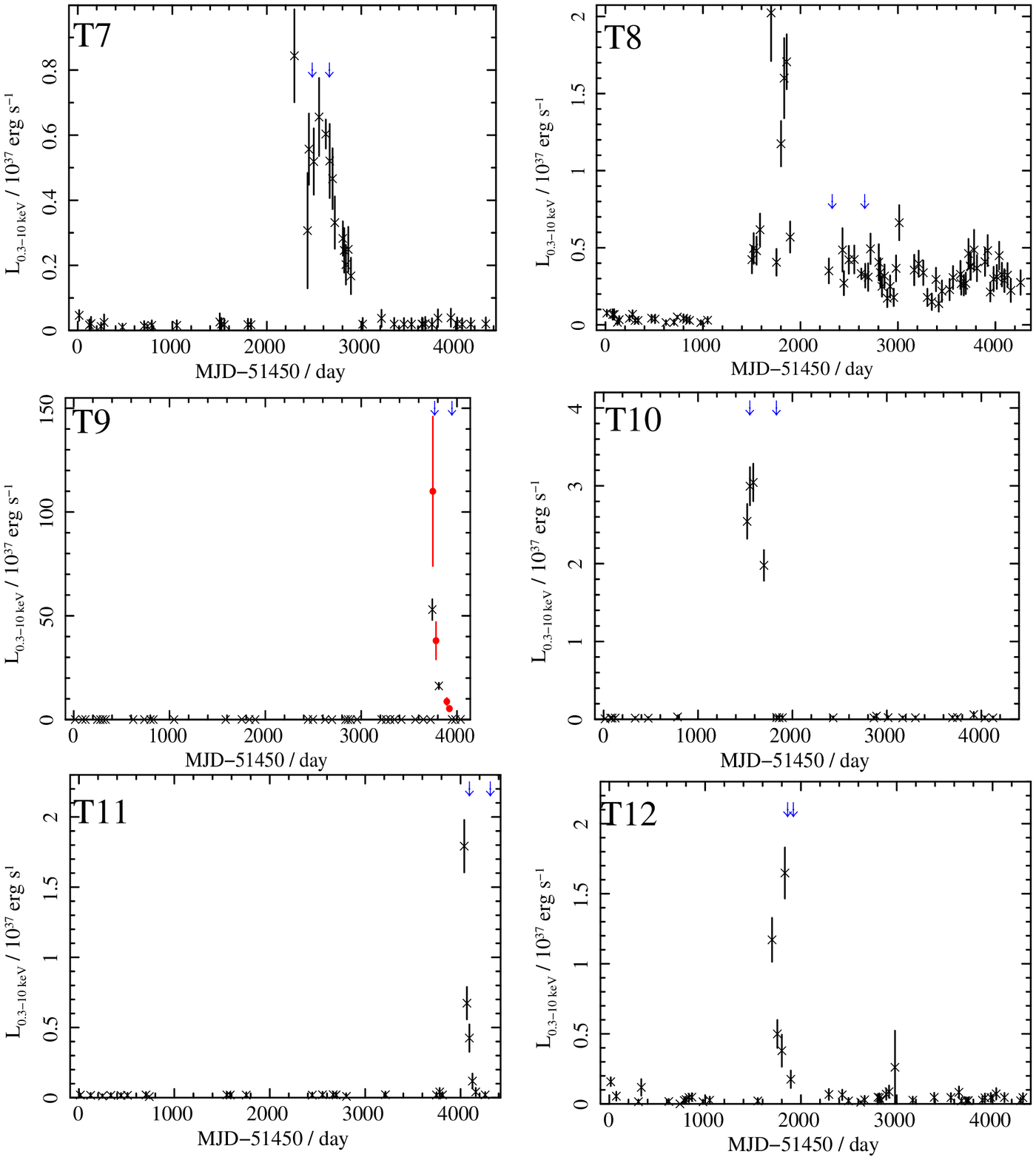}
\caption{continued}
\end{figure*}

\begin{figure*}
\epsscale{1.85}
\plotone{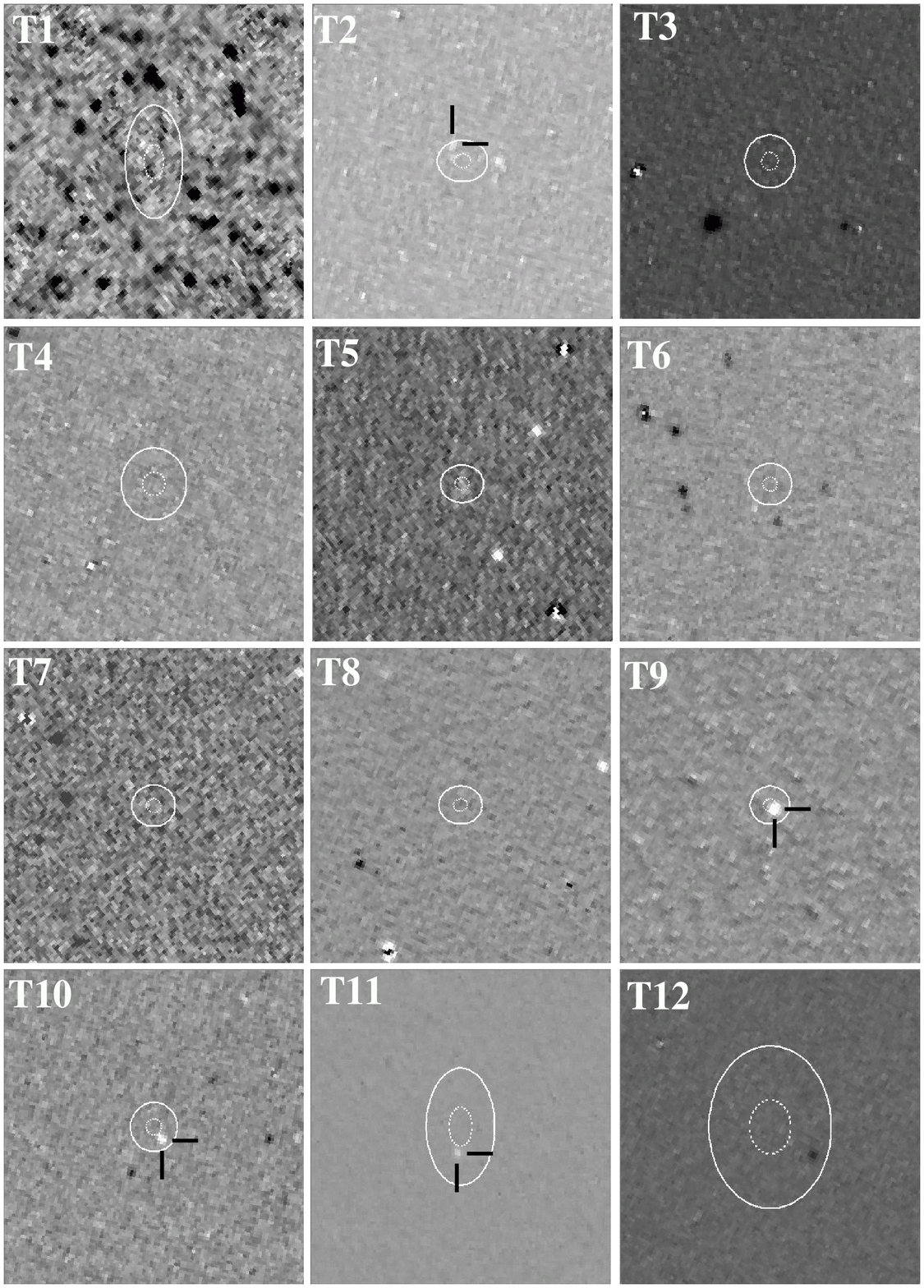}
\caption{Difference images for the 12 transients, obtained by subtracting the HST off image from the HST on image. Each image is 5$''$$\times$5$''$.  White is positive, and black is negative. The dotted ellipses indicate 1$\sigma$ uncertainties, while the solid ellipses indicate 3$\sigma$ uncertainties in the location of the X-ray source. We subtracted a V-Band off image from a B-band on image for T1;  there are many black stars because they are brighter in V than in B. \label{difim}}
\end{figure*}

\begin{figure*}
\epsscale{1.85}
\plotone{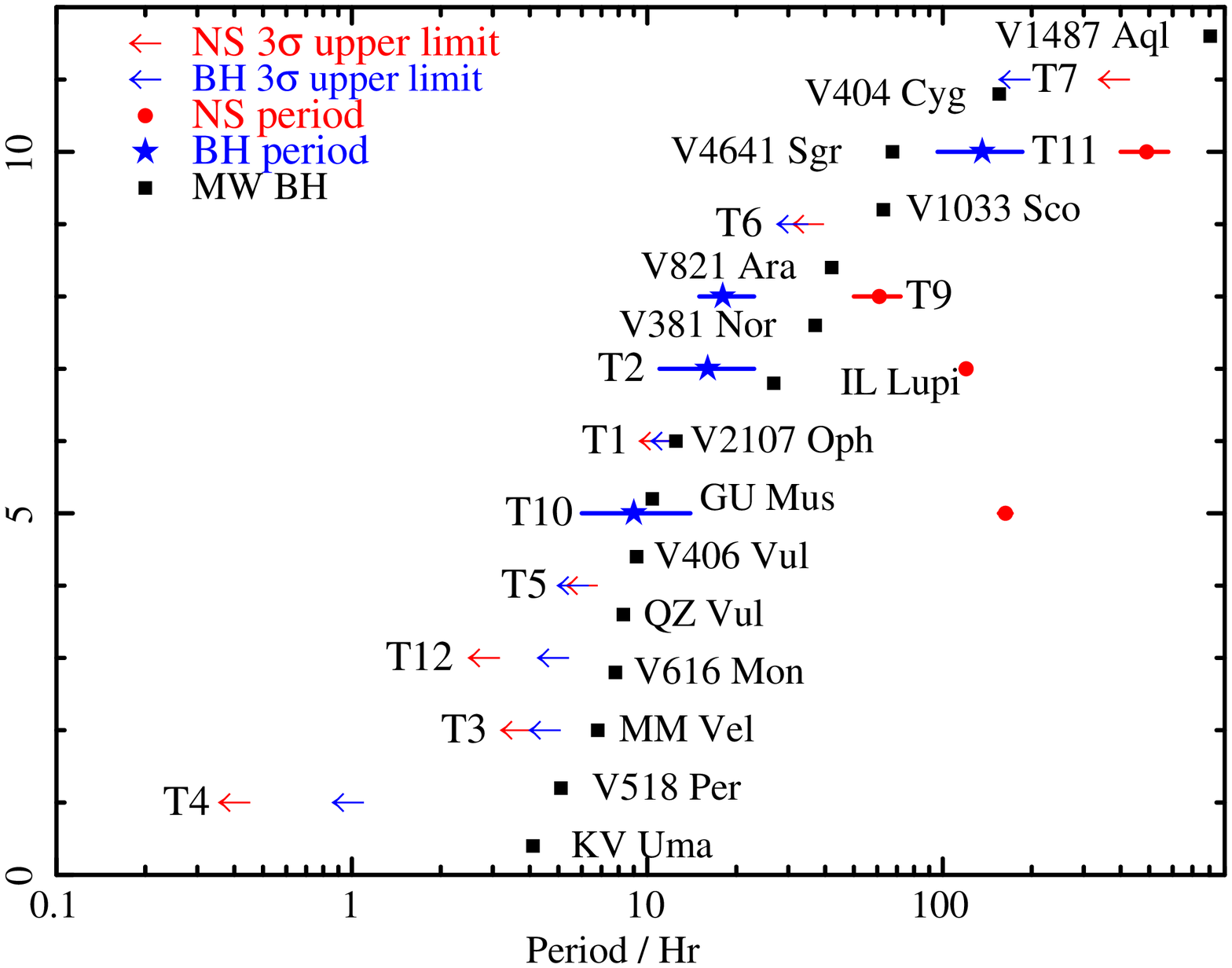}
\caption{Period distributions of the twelve transients, assuming the VPM94 relation for a NS primary and the M12 relation for a BH primary; we have ordered the transients by increasing BH period.  Circles indicate NS periods, stars indicate BH periods, and arrows indicate 3$\sigma$ upper limits. We also show the period distribution for Milky Way BH binaries for comparison, represented by squares.  \label{perdist}}
\end{figure*}

\begin{figure*}
\epsscale{1.85}
\plotone{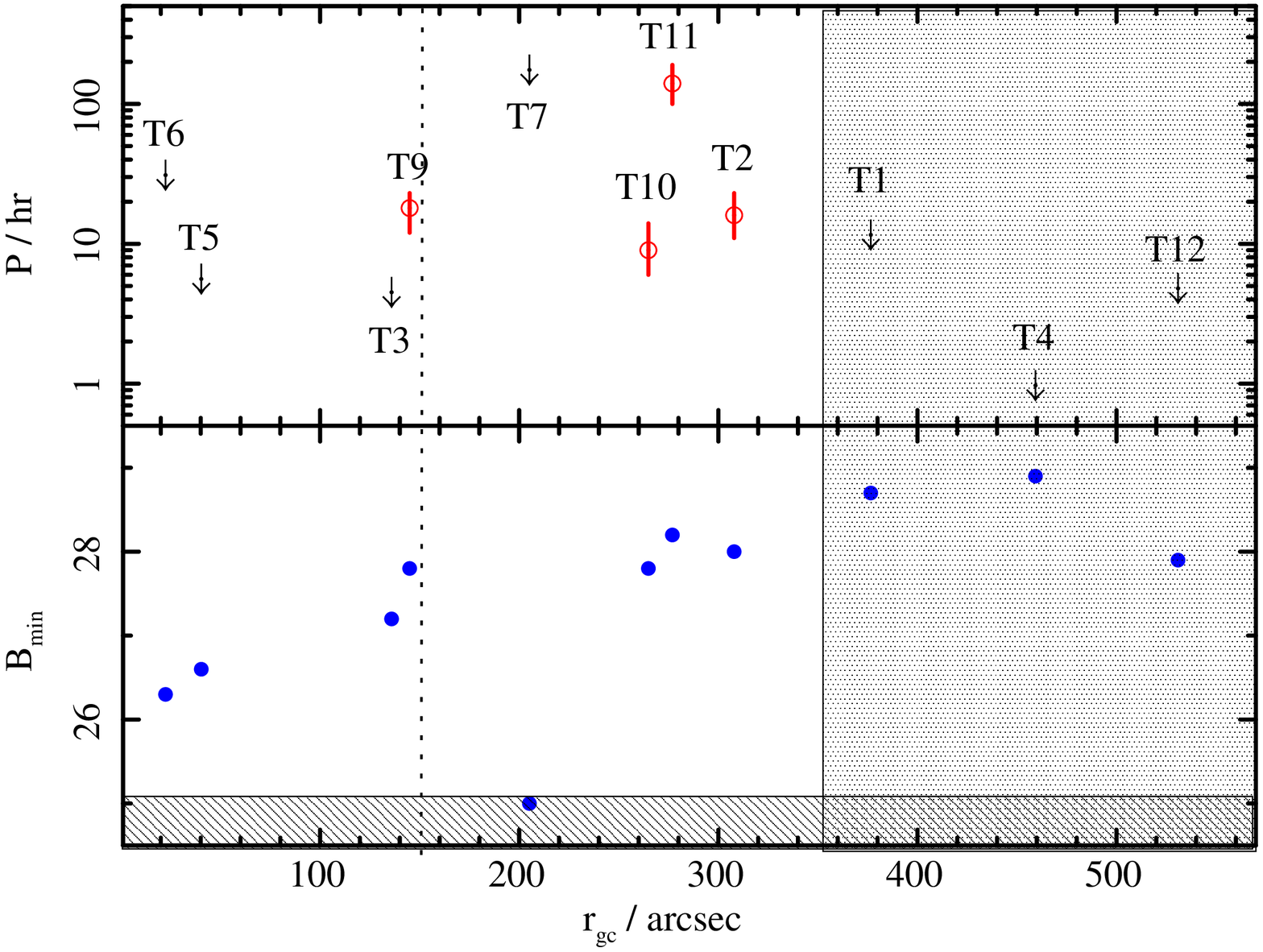}
\caption{{\it Top Panel}: Period vs. galacto-centric radius; we compare the periods of transients within 350$''$ of the nucleus, and the dot-shaded region is outside this sample. {\it Bottom Panel}: B band sensitivity limit vs galacto-centric radius; the diagonally  shaded region covers sources that exhibited no X-ray variation between their HST on and off observations. The sensitivity increase by 2.5 magnitudes (i.e. a factor of 10). } \label{rvp}
\end{figure*}

\clearpage

\begin{table*}
\begin{center}
\caption{ Journal of HST ACS observations. For each observation we provide the HST ID, date, exposure time in seconds, RA and Dec of pointing, and target. We also show the closest Chandra ACIS observation, and the time of the Chandra observation relative to the HST observation.} \label{journ}
\begin{tabular}{cccccccccccc}
\tableline\tableline
Obs & ID &  Date & Exp& RA & Dec & Target & Chan Obs & Sep. \\
\tableline 
1 & j8vp02010 & 2003-12-25 & 2200 &  00:42:56.03 &  +41:12:19.0 & T10 on &  4680 & +2 days  \\ 
2 & j8vp03010 & 2004-01-23 & 2200 &  00:42:43.86 &  +41:16:30.1 & T6 on &  4681 & $+$8 days \\ 
3 & j8vp04010 & 2004-06-14 & 2200 &  00:42:33.39 &  +41:17:43.4 & T3 on  &  4682 & $-$22 days \\ 
4 & j8vp05010 & 2004-08-15 & 2200 &  00:42:41.50 &  +41:17:00.0 & T5 on  &  4719 & +29 days \\ 
5 & j8vp07010 & 2004-10-02 & 2200 &  00:42:56.04 &  +41:12:19.1 & T10 off&  4721 & +2 days\\ 
6 & j8vp08010 & 2004-11-02 & 2200 &  00:43:09.93 &  +41:23:32.6 & T12 on &  4721 & $-$29 days \\ 
7 & j8vp09010 & 2004-11-22 & 2200 &  00:42:41.49 &  +41:17:05.2 & T5 off&  4723 &  +13 days \\ 
8 & j8vp10010 & 2005-01-01 & 2200 &  00:43:09.91 &  +41:23:32.5 & T12 off &  4723 &  $-$27 days \\ 
9 & j9ju01010 & 2006-02-10 & 4360 &  00:42:52.38 &  +41:16:49.2 & T8 on &  7136 & $-$35 days \\ 
10 & j9ju02010 & 2006-07-09 & 4360 &  00:42:33.89 &  +41:23:31.1 & T4 on&  7139 & +22 days \\ 
11 & j9ju03010 & 2006-07-15 & 4352 &  00:42:51.17 &  +41:19:17.9 & T7 on &  7139 & $-$16 days \\ 
12 & j9ju04010 & 2006-08-27 & 4344 &  00:42:16.08 &  +41:19:26.1 & T1 on  &  7139 & $-$27 days \\ 
13 & j9ju05010 & 2006-10-25 & 2148 &  00:42:17.34 &  +41:15:37.0 & T2 on  &  7140 & $-$32 days \\ 
14 & j9ju06010 & 2007-01-10 & 4672 &  00:42:52.38 &  +41:16:49.2 & T8 off &  8183 & +4 days \\ 
15 & j9ju07010 & 2007-01-12 & 4672 &  00:42:33.88 &  +41:23:31.2 & T4 off&  8183 &  +2 days \\ 
16 & j9ju08010 & 2007-01-12 & 4672 &  00:42:51.14 &  +41:19:18.0 & T7 off&  8183 &  +2 days \\ 
17 & j9ud17010 &  2009-08-25 & 4360 &  00:42:17.31 &  +41:15:37.5 & T2 off&  10555 &  $-$55 days \\ 
18 & jb9d15010 & 2010-01-21 & 4360 &  00:42:53.15 &  +41:14:22.9 & T9 on &  11278 & +14 days \\ 
19 & jb9d20010 &  2010-07-20 & 4360 & 00:42:53.15 &  +41:14:22.9 & T9 off&  11840 &  $-$3 days  \\ 
20 & jbdl01010 & 2010-12-09 & 5472 &  00:42:57.00 &  +41:20:05.0 & T11 on &  12162 & $-$1 day \\ 
21 & j9ju08010 & 2011-07-20 & 4672 &  00:42:51.17 & +41:10:17.9 & T11 off & 12972 & $-$5 days \\
\tableline
\end{tabular}
\end{center}
\end{table*}

\clearpage

\begin{table*}
\begin{center}
\caption{For each object, we provide the X-ray position, with 1$\sigma$ uncertainties, along with the 0.3--10 keV luminosity during the closest Chandra observation to the HST on observation, and the estimated 2--10 keV luminosity at the time of the HST on observation; the luminosities are normalized to 10$^{37}$ erg s$^{-1}$. We then provide the B band apparent Vega magnitude, and the derived absolute V band magnitude. We finally estimate the orbital period using the VPM94 method ($P_{\rm NS}$), and M12 method ($P_{\rm BH}$).  } \label{translist}
\renewcommand{\tabcolsep}{3pt}
\begin{tabular}{cccccccccccc}
\tableline\tableline
Targ & RA & $\sigma_{\rm RA}$ & Dec & $\sigma_{\rm Dec}$ & $L_{\rm Ob}$ & $L_{\rm Est}$ & $B$ & $M_{\rm V}$ & $P_{\rm NS}$  & $P_{\rm BH}$  \\
\tableline 
1 & 00:42:16.106 & 0.015 & +41:19:26.36 & 0.27 & 8.9$\pm$2.3 & 0.3 & $>$28.7 & 2.45 & $<$10.6 & $<$11.6\\
2 & 00:42:17.330 & 0.012 & +41:15:37.48 & 0.11 & 12.8$\pm$1.2 & 2.89 & 24.7$\pm$0.3 & $-$0.2 & 120$\pm$5 & 16$^{+7}_{-5}$\\
3 & 00:42:33.426 & 0.012 & +41:17:03.56 & 0.14 & 3.7$\pm$0.3 & 1.59 & $>$27.2 & 2.3 & $<$3.6 & $<$4.5\\
4 & 00:42:33.883 & 0.016 & +41:23:31.49 & 0.19 & 2$\pm$0.2 & 0.88 & $>$28.9 & 4 & $<$0.4 & $<$0.97\\
5 & 00:42:41.805 & 0.010 & +41:16:35.94 & 0.10 & 5.8$\pm$0.8 & 2.52 & $>$26.6 & 1.7 & $<$6.0 & $<$5.6\\
6 & 00:42:43.834 & 0.011 & +41:16:30.60 & 0.11 & 0.94$\pm$0.12 & 0.53 & $>$26.3 & 1.4 & $<$35 & $<$31\\
7 & 00:42:51.199 & 0.012 & +41:19:17.81 & 0.13 & 0.51$\pm$0.1 & 0.35 & $>$25 & 2.9 & $<$380 & $<$175\\
8 & 00:42:52.433 & 0.010 & +41:16:48.75 & 0.10 & 0.35$\pm$0.08 & 0.29 &$ >$23.9 & $-$1 & $<$2400 & $<$1000\\
9 & 00:42:53.181 & 0.010 & +41:14:22.83 & 0.10 & 38$\pm$9 & 18.62 & 23.91$\pm$0.08 & $-$0.99 & 61$\pm$11 & 18$^{+5}_{-6}$\\
10 & 00:42:56.049 & 0.012 & +41:12:18.64 & 0.13 & 3$\pm$0.2 & 1.77 & 24.52$\pm$0.02 & $-$0.38 & 163$\pm$8 & 9$^{+5}_{-3}$\\
11 & 00:42:56.991 & 0.017 & +41:20:05.35 & 0.31 & 0.39$\pm$0.10 & 0.2 & 24.87$\pm$0.09 & $-$0.03 & 490$\pm$90 & 140$^{+50}_{-40}$\\
12 & 00:43:09.963 & 0.030 & +41:23:32.56 & 0.43 & 1.65$\pm$0.12 & 0.53 &$>$ 27.9 & 3 & $<$2.8 & $<$4.8\\
\tableline 
\end{tabular}
\end{center}
\end{table*}

\clearpage

\begin{table*}
\begin{center}
\caption{Spectral fit parameters for transients with $>$200 net source counts in their spectra, assuming an absorbed disk blackbody model. For each source we give the net source counts, absorption, inner disk temperature and $\chi^2$/dof for the best fits.  We also provide the ratio of luminosities in the 2--10 and 0.3--10 keV bands. We note that the T9 spectrum is piled up; this is accounted for in Noorae et al. (2012)} \label{spectab}
\begin{tabular}{cccccccccccc}
\tableline\tableline
Object & Net source counts & $N_{\rm H}$ / 10$^{21}$ atom cm$^{-2}$ & k$T$ / keV & $\chi^2$/dof & $L_{2-10}/L{0.3-10}$\\
\tableline 
T1 & 374 & 2.5$\pm$1.9 & 1.5$\pm$0.3 & 10/15 & 0.55 \\
T2 & 756 & 0.67 & 0.54$\pm$0.04 & 35/31 &0.25 \\
T5 & 330 & 0.67 & 1.1$\pm$0.2 & 4/12 & 0.59\\
T9 & 1135 & 0.67 & 0.8 & 64/49& 37\\
\tableline
\end{tabular}
\end{center}
\end{table*}

\end{document}